\documentclass[%
 reprint,
 amsmath,amssymb,
 aps,
]{revtex4-2}

\usepackage{graphicx}
\usepackage{dcolumn}
\usepackage{bm}

\usepackage{multirow}
\usepackage{titlesec}
\usepackage{orcidlink}

\usepackage[T1]{fontenc}
\usepackage[utf8]{inputenc}
\usepackage{mathtools}
\newcommand\myeq{\mathrel{\overset{\makebox[0pt]{\mbox{\normalfont\scriptsize\sffamily 0}}}{=}}}

\begin{document}

\title{Effects of the matter Lagrangian degeneracy in $f(Q,T)$ gravity}

\author{José Antonio Nájera \orcidlink{0000-0001-9738-7704}}
\email{najera111@ciencias.unam.mx}
\author{and Carlos Aráoz Alvarado
\orcidlink{0000-0002-7095-7874}} 
 \email{carlosaraoz@ciencias.unam.mx}
\affiliation{%
 Facultad de Ciencias, Universidad Nacional Autónoma de México, Investigación Científica, C.U., Coyoacán, Ciudad de México 04510, México
}%

\date{\today}

\begin{abstract}
In this paper,  we investigated the theoretical and cosmological effects of the matter Lagrangian degeneracy in  an extension of the Symmetric Teleparallel Equivalent of General Relativity, denoted as $f (Q, T )$ gravity. This degeneracy comes from the fact that both $\mathcal{L}_m = p$ and $\mathcal{L}_m = -\rho$ can give rise to the stress-energy tensor of a perfect fluid. The $f(Q,T)$ equations depend on the form of the matter Lagrangian and hence they also have a degeneracy that has influence on the density evolution of dust matter and radiation, on the form of the generalized Friedmann equations and also shows changes in the cosmological parameters confidence regions when performing Monte Carlo Markov Chains analyses. Our results suggest that since changing the matter Lagrangian causes different theoretical and cosmological results, future studies in $f(Q,T)$ gravity should consider both forms of the matter Lagrangians to account for the different mathematical and observational results. 
\end{abstract}

\maketitle


\section{Introduction}

In the standard theory of General Relativity, gravity is a geometric entity caused by a curvature tensor called the Riemann tensor $R^\alpha_{\;\;\beta\mu\nu}$. Inside the framework of this theory, the $\Lambda$CDM model has been formulated and strongly tested. Even though this model is able to describe the accelerated expansion of the Universe, it has several problems yet to be solved. One of them is the fact that the nature of the exotic fluid called \textit{dark energy} (the one that causes the accelerated expansion of the universe) is still unknown. Another problem that has been growing in recent years with the acquisition of more data and reduced systematic errors is called \textit{the Hubble constant tension problem}. It has been shown that there exists a $5\sigma$ tension \cite{riess2021comprehensive} between the value of the Hubble constant measured by SH0ES \cite{riess2021comprehensive} and the one from the Planck collaboration \cite{aghanim2020planck}. The cause of this problem is still unknown, however, this tension suggests that new physics beyond the standard $\Lambda$CDM model is required. 

One possibility to solve these problems is to modify the theory of gravity. This can be done by changing the Ricci scalar $R$ in the Einstein-Hilbert Lagrangian to an arbitrary function of the Ricci scalar. These theories are called $f(R)$ gravity \cite{sotiriou2010f}. However, in the metric approach, they also work in the General Relativity framework, with gravity being driven by curvature while the torsion and non-metricity are equal to zero, and the connection is equal to the Levi-Civita one. In addition to this, in this approach, a pseudo-Riemannian spacetime is required. Nevertheless, it is important to mention that $f(R)$ gravity can also be formulated with metricity in the Palatini/metric-affine approach, and also in the hybrid formulation.

Another set of theories that are constructed by varying the Einstein-Hilbert action and that modify the theory of gravity are known as nonlocal theories of General Relativity. Nonlocally modified models of gravity, motivated by quantum loop corrections, have been investigated as a device for explaining current cosmic acceleration \cite{deser2007nonlocal}. Dark energy has been investigated   under this formalism by adding a term $m^2 R \; \square^{-2} R$ to the Einstein-Hilbert action \cite{maggiore2014nonlocal}. The thermodynamic properties of clusters of galaxies have been studied under this nonlocal formalism  by adding a distortion function $f(\square^{-1}R)$ of the nonlocal parameter $\square^{-1}R$ to the Einstein-Hilbert action and considering the weak-field limit approximation 
\cite{araoz}. In all of these cases, we are still rooted under the framework of General Relativity. Therefore, if a simpler formulation is required, the way to build gravity needs to be changed. 

One form to study gravity without the need of the General Relativity framework, and therefore the necessity of a pseudo-Riemannian spacetime, is to assume that curvature is equal to zero and hence the Riemann tensor is zero ($R^\alpha_{\;\;\beta\mu\nu}=0$). By setting curvature to zero, we enter in the so-called \textit{teleparallel formulation} \cite{maluf2013teleparallel, bahamonde2021teleparallel}. In this framework, the Weitzenböck connection can be built $\Gamma^\alpha_{\mu\nu} = e^{a\alpha} \partial_\mu e_{a\nu}$. At this point, the connection can be restricted in two ways. The first one is to set the covariant derivative of the metric, the \textit{non-metricity} to zero $Q_{\alpha \mu \nu} = \nabla_\alpha g_{\mu\nu} = 0$ and letting gravity be driven by torsion $T^\alpha_{\;\;\mu \nu} = 2\Gamma^\alpha_{\;\;[\mu\nu]} \neq 0$ \cite{bahamonde2021teleparallel}. This foundation is called \textit{metric teleparallel gravity}. The second form to restrict the connection is to set the torsion to zero $T^\alpha_{\;\;\mu \nu} = 2\Gamma^\alpha_{\;\;[\mu\nu]} = 0$ and letting gravity be driven by the non-metricity tensor $Q_{\alpha \mu \nu} = \nabla_\alpha g_{\mu\nu} \neq 0$. This case is also part of the teleparallel formulation, but since the connection is symmetric in its lower indices, it is called \textit{symmetric teleparallel gravity} \cite{nester1998symmetric,adak2006lagrange}. 

Inside the symmetric teleparallel formulation, it is possible to build an equivalent formulation to General Relativity, denoted as the Symmetric Teleparallel Equivalent of General Relativity (STEGR) \cite{jimenez2018coincident} with an action $S = \int \sqrt{-g} (-\kappa^2 Q + \mathcal{L}_m)$ with $\kappa^2 = 8\pi G$ and $Q$ the non-metricity scalar (a scalar derived from contractions of the non-metricity tensor). The geometrical framework for this formulation of General Relativity is simple because there is no curvature nor torsion and the non-metricity scalar $Q$ is rooted as the fundamental geometrical object that describes the gravitational interaction \cite{beltran2019geometrical}. The present paper will focus in this third form to build gravity.

In a similar way as in General Relativity, we can consider an extension of the STEGR by taking an arbitrary function of the non-metricity scalar which gave birth to $f(Q)$ gravity \cite{jimenez2018coincident}. These theories can have the potential to explain the accelerated expansion of the Universe with a geometrical nature instead of a exotic perfect fluid with negative pressure. The cosmology and linear perturbations of $f(Q)$ gravity where studied in \cite{jimenez2020cosmology} and, it has been shown that some functions within these theories can challenge the standard $\Lambda$CDM model \cite{anagnostopoulos2021first,atayde2021can}, putting them in a great position to make further studies and determine whether they can solve the cosmological constant and Hubble constant tension problems. 

Another extension of the STEGR gravity has been considered   in the framework of the metric affine formalism \cite{harko2018coupling}, where the metric and the affine connection are regarded as independent variables, by introducing a theory where the non-metricity $Q$ is non-minimally coupled to the matter Lagrangian. This non-minimal coupling calls for the non-conservation of the energy-momentum tensor, and consequently an extra force appears in the geodesic equation of motion.

In regards to non-minimally coupling between geometry and matter, a theory called $f(Q,T)$ gravity was proposed \cite{xu2019f}, based on the non-minimal coupling between the non-metricity $Q$ and the trace $T$ of the matter stress-energy tensor. Similarly to the standard curvature-trace of energy-momentum tensor couplings, in the $f(Q, T )$ theory, the coupling between $Q$
and $T$ leads to the non-conservation of the energy-momentum tensor.

$f(Q,T)$ gravity has been deeply studied recently. The Chaplygin gas in the context of $f(Q,T)$ was explored in \cite{gadbail2022generalized}, observational constraints of $f(Q,T)$ gravity functions are presented in \cite{singh2022cosmological,sokoliuk2022cosmological,gadbail2021power,gadbail2021viscous,agrawal2021matter,pati2021model,arora2021constraining,arora2021constraining,godani2021frw,zia2021transit,arora2020energy,arora2021constraining,arora2020f} and the linear theory of cosmological perturbations were explored in \cite{najera2022cosmological}. In addition to this, it has been shown that a particular quadratic function on $T$ ($f (Q, T ) = -(Q + 2\Lambda)/G - ((16\pi)^2 G b)/(120 H_0^2)T^2$) can challenge the $\Lambda$CDM standard model, showing a substantial preference against it \cite{najera2021fitting} with Supernova and cosmic clocks data. However, until now, this theory still needs more observational studies, for instance, whether the theory can pass the solar system tests naturally or if it needs some constraints.

However, when studying theoretical and observational aspects of $f(Q,T)$ gravity, the standard way is to take $\mathcal{L}_m = p$ as the matter Lagrangian that gives rise to the stress-energy tensor of a perfect fluid. Since, as we will see, there is another form of the matter Lagrangian $\mathcal{L}_m = -\rho$, this induces a degeneracy in the equations of $f(Q,T)$ gravity. Moreover, this degeneracy impacts the field equations of all theories includying a non-minimally coupling of the trace of the stress-energy tensor $T$ with geometry, for example, $f(R,T)$ theory \cite{carvalho2021general}. In this paper, we will study the theoretical and cosmological effects that this degeneracy has in $f(Q,T)$ gravity. We will study how the generalized Friedmann equations, the density evolution of dust matter and radiation change when considering a different Lagrangian. Furthermore, we will perform a Numerical analysis for some cosmological models to explore even more the effects of the matter Lagrangian degeneracy when doing Monte Carlo Markov Chains analyses.

The paper is divided as follows: 
in Sec.~\ref{sec: stress-energy tensor of a perfect fluid 2}, we discussed the two forms of the matter Lagrangian to get the stress-energy tensor of a perfect fluid. In Sec.~\ref{sec:General $f(Q,T)$ gravity 3}, we presented the general theory under $f(Q,T)$ gravity. In Sec.~\ref{sec:coincident Gauge 4}, we reviewed the coincident gauge. In Sec.~\ref{sec:Effects  in Friedmann equations 5}, we derived the generalized Friedmann equations and explored the effects of the matter Lagrangian degeneracy in them. In Sec.~\ref{sec:Effects 6}, we obtained the density evolution and the effects of the matter Lagrangian degeneracy.  In order to see the cosmological effects that a different matter Lagrangian has, in Sec.~\ref{sec:Specific Cosmological Models 7} we considered some cosmological models and obtained their generalized Friedmann equations and density evolution of matter and radiation for each model. We also performed MCMC analyses for each one. Finally, in Sec.~\ref{sec:Conclusions}, we gathered our main results and findings.  

\section{The stress-energy tensor of a perfect fluid}
\label{sec: stress-energy tensor of a perfect fluid 2}

It is standard in cosmology to assume that the Universe is composed by a perfect fluid and therefore its stress-energy tensor is given by
\begin{equation}
    \label{eqn:stress-energyPerfectFluid}
    T_{\mu \nu} = (\rho + p) u_\mu u_\nu + p g_{\mu \nu}.
\end{equation}

This tensor is also defined in terms of variations as
\begin{equation}
    \label{eqn:stress-energyGeneral}
    T_{\mu \nu} = - \frac{2}{\sqrt{-g}} \frac{\delta (\sqrt{-g} \mathcal{L}_m)}{\delta g^{\mu \nu}} = g_{\mu \nu} \mathcal{L}_m - 2 \frac{\delta \mathcal{L}_m}{\delta g^{\mu \nu}},
\end{equation}
where $\mathcal{L}_m$ is the matter Lagrangian and $g=det(g_{\mu\nu})$. We can see that with this definition of the stress-energy tensor and Eq.~(\ref{eqn:stress-energyPerfectFluid})  if we assume that the matter Lagrangian is given as a function of a real scalar field $\phi$ and $X$ (such that $ \mathcal{L}_m = \mathcal{L}_m(\phi, X)$) where $X = -\frac{1}{2} \nabla^\mu \phi \nabla_\mu \phi$, then the matter Lagrangian is given by  $\mathcal{L}_m = p$ \cite{avelino2018perfect}, where $p$ is the pressure. 

However, there is an alternative form of the matter Lagrangian $\mathcal{L}_m$ that gives rise to the perfect fluid stress-energy tensor Eq.~(\ref{eqn:stress-energyPerfectFluid}). This form is given by  $\mathcal{L}_m = -\rho$ \cite{mendoza2021matter}, where $\rho$ is the density. Therefore, we have two forms of the matter Lagrangian that give the same perfect fluid stress-energy tensor. 

\section{General $f(Q,T)$ gravity}
\label{sec:General $f(Q,T)$ gravity 3}

Since $f(Q,T)$ is an extension of the symmetric teleparallel formalism, gravity is driven by non-metricity instead of curvature as in General Relativity. 

The field equations of $f(Q,T)$ gravity come from its general action, which is given by 
\begin{equation}
    \label{eqn:fQTaction}
    S = \int d^4x \sqrt{-g} \left( \frac{1}{16\pi} f(Q,T) + \mathcal{L}_m \right),
\end{equation}
where $Q$ is the non-metricity scalar, and $T$ is the trace of the stress-energy tensor. The non-metricity tensor is defined as the covariant derivative of the metric
\begin{equation}
    Q_{\alpha \mu \nu} \equiv \nabla_\alpha g_{\mu \nu}.
    \label{eqn:non-metricity tensor}
\end{equation}

The so-called non-metricity scalar is given in terms of a linear combination of four possible contractions of the non-metricity tensor that give a scalar \cite{xu2019f}
\begin{align}
    Q = -\frac{1}{4} \biggl(& -Q^{\alpha \mu \nu} Q_{\alpha \mu \nu} + 2 Q^{\alpha \mu \nu} Q_{\nu \alpha \mu} \nonumber \\
    &-2Q^\alpha \tilde{Q}_{\alpha} + Q^\alpha Q_\alpha  \biggr),
    \label{eqn:Q}
\end{align}
where $Q_{\alpha} = g^{\mu \nu} Q_{\alpha \mu \nu}$ and $\tilde{Q}_{\alpha} = g^{\mu \nu} Q_{\nu \alpha \mu}$. It is also useful to define the deformation tensor \cite{xu2019f}
\begin{equation}
    L^{\alpha}_{\;\;\mu\nu} = -\frac{1}{2} g^{\alpha \beta} \left( Q_{\nu \mu \beta} + Q_{\mu \beta \nu} - Q_{\beta \mu \nu} \right),
    \label{eqn:deformation tensor}
\end{equation}
and the superpotential \cite{xu2019f}
\begin{equation}
    P^\alpha_{\;\;\mu\nu} = -\frac{1}{2} L^\alpha_{\;\;\mu\nu} + \frac{1}{4} \left( Q^\alpha - \tilde{Q}^\alpha \right) g_{\mu \nu} - \frac{1}{4} \delta^\alpha_{\;\;(\mu} Q_{\nu)}.
\end{equation}

By varying the action Eq.~(\ref{eqn:fQTaction}) with respect to the metric and according to the principle of stationary action (which makes the variation of the action with respect to the metric zero) we get the field equations \cite{xu2019f}
\begin{align}
    -\frac{2}{\sqrt{-g}} &\nabla_\alpha \left( f_Q \sqrt{-g} P^\alpha_{\;\;\mu\nu} \right) - \frac{1}{2} f g_{\mu \nu} + f_T (T_{\mu \nu} + \Theta_{\mu \nu}) \nonumber \\
    &-f_Q (P_{\mu \alpha \beta} Q_{\nu}^{\;\;\alpha\beta} - 2 Q^{\alpha\beta}_{\;\;\;\;\mu}P_{\alpha \beta \nu}) = 8\pi T_{\mu\nu},
\end{align}
where $f_Q \equiv df/dQ$, $f_T = df/dT$, $T_{\mu \nu}$ is the stress-energy tensor and $\Theta_{\mu \nu} = g^{\alpha \beta} \dfrac{\delta T_{\alpha \beta}}{\delta g^{\mu \nu}}$. By varying the action Eq.~(\ref{eqn:fQTaction}) with respect to the connection instead of the metric we get the connection field equations
\begin{equation}
\label{eqn:connectionFieldEquations}
    \nabla_\mu \nabla_\nu \left( \sqrt{-g} f_Q P^{\mu \nu}_{\;\;\;\;\alpha} + 4\pi H_{\alpha}^{\;\;\mu\nu} \right) = 0,
\end{equation}
with $H_{\alpha}^{\;\;\mu\nu}$ the hypermomentum tensor density \cite{xu2019f}.
\begin{equation}
    H_{\alpha}^{\;\;\mu \nu} = \frac{\sqrt{-g}}{16\pi} f_T \frac{\delta T}{\delta \Gamma^\alpha_{\;\;\mu\nu}} + \frac{\delta(\sqrt{-g} \mathcal{L}_m)}{\delta \Gamma^\alpha_{\;\;\mu\nu}}.
\end{equation}

Raising an index to the field equations
\begin{align}
\label{eqn:raisedFieldEquations}
    -\frac{2}{\sqrt{-g}} &\nabla_\alpha(f_Q \sqrt{-g} P^{\alpha \mu}_{\;\;\;\;\nu}) -\frac{1}{2} f \delta^\mu_{\;\;\nu} + f_T (T^\mu_{\;\;\nu} + \Theta^\mu_{\;\;\nu}) \nonumber \\
    & -f_Q P^{\mu\alpha\beta} Q_{\nu\alpha\beta} = 8\pi T^\mu_{\;\;\nu}.
\end{align}

\section{Coincident Gauge}
\label{sec:coincident Gauge 4}

In order to compute the covariant derivatives, we need to define the connection. As aforesaid, in the symmetric teleparallel formalism, both the curvature and torsion are equal to zero and gravity is driven by non-metricity. Then, in this framework
\begin{equation}
\label{eqn:STGPostulates}
\begin{split}
    R^{\alpha}_{\;\;\beta\mu\nu} = 0, \\
    T^{\mu}_{\;\;\alpha\beta} = 0, \\
    Q_{\alpha\mu\nu} \neq 0,
\end{split}
\end{equation}
where $R^{\alpha}_{\;\;\beta\mu\nu}$ is the Riemman tensor, $T^{\mu}_{\;\;\alpha\beta}$ the torsion tensor and $Q_{\alpha\mu\nu}$ the non-metricity tensor. The most general connection that fulfills the three conditions Eq.~(\ref{eqn:STGPostulates}) is given by \cite{beltran2019geometrical,runkla2018family}
\begin{equation}
    \Gamma^\alpha_{\;\;\mu\nu} = \frac{\partial x^\alpha}{\partial \xi^\beta} \partial_\mu \partial_\nu \xi^\beta,
\end{equation}
where $\xi^\beta$ are a set of functions. It is standard to take the so-called \textit{coincident} gauge in this symmetric teleparallel framework. In this gauge, $\xi^\alpha = x^\alpha$ and hence the connection is zero. However, when taking this particular gauge, we are only studying a very restricted class of geometries \cite{hohmann2021general}. Two recent papers in the literature have derived the most general form of the connection that satisfies zero curvature, zero torsion and a FLRW metric \cite{hohmann2021general,d2021revisiting}. These results can explore the full set of geometries compatible with symmetric teleparallelism. The study of $f(Q,T)$ gravity with that general connection is beyond the scope of this paper but it constitutes a possible topic of study in future works. \\

In this gauge, the deformation tensor can be written as minus the Levi-Civita connection as we will see now. Einstein proposed \cite{einstein1922hamiltonsches} the following Lagrangian
formulation for his field equations
\begin{equation}
    L_E=g^{\mu \nu}\Big(\big\{^\alpha _{\;\;\beta\mu} \big\} \big\{^\beta _{\;\;\nu\alpha} \big\} -\big\{^\alpha _{\;\;\beta\alpha} \big\} \big\{^\beta _{\;\;\mu \nu} \big\}\Big)
    \label{eqn:LE}
\end{equation}

presenting the Levi-Civita connection, (here as the Christoffel symbols), of the metric $g_{\mu \nu}$

\begin{equation}
    \big\{^\lambda _{\;\;\mu \nu} \big\}=\frac{1}{2}g^{\lambda \beta}(\partial_\mu g_{\beta \nu}+\partial_\nu g_{\beta \mu}-\partial_\beta g_{\mu \nu})
    \label{eqn:Christoffel symbols}
\end{equation}
Since the Lagrangian Eq.~(\ref{eqn:LE}) is not covariant, we can foster the partial derivatives of the metric in Eq.~(\ref{eqn:Christoffel symbols}) to covariant ones in order to foresee it. Introducing an independent “Palatini connection” $\Gamma^\alpha_{\mu \nu}$, with a covariant derivative $\nabla_\alpha$ \cite{harko2018coupling}, so we can define the tensor 
\begin{equation}
    L^{\alpha}_{\;\;\mu\nu} = -\frac{1}{2} g^{\alpha \beta} \left( \nabla_{\nu} g_{\mu \beta} + \nabla_{\mu} g_{ \beta \nu} - \nabla_{\beta} g_{ \mu \nu} \right),
\end{equation}

which is the deformation tensor Eq.~(\ref{eqn:deformation tensor}) using the definition of the non-metricity tensor Eq.~(\ref{eqn:non-metricity tensor}). If the covariant derivative reduces to the partial one, the invariant
\begin{equation}
    Q=-g^{\mu \nu}(L^\alpha_{\;\;\beta \mu}L^\beta_{\;\;\nu \alpha}-L^\alpha_{\;\;\beta \alpha}L^\beta_{\;\;\mu \nu})
\end{equation}
is equivalent to minus the Einstein Lagrangian (-$L_E$), ad-hoc,
\begin{equation}
    \nabla_\alpha \myeq \partial_\alpha,\;\;\;\; Q\myeq -L_E
\end{equation}

This gauge choice, designated with the $0$, is known as the coincident gauge. We will work with the coincident gauge in this paper to see the effects that the matter Lagrangian degeneracy has on the generalized Friedmann equations and on the stress-energy evolution in this particular gauge. 

\section{Effects of the matter Lagrangian degeneracy in the generalized Friedmann equations}
\label{sec:Effects  in Friedmann equations 5}

By working in the coincident gauge, we need to set the connection to zero and therefore the covariant derivatives are standard derivatives. We will also work in an isotropic, spatially flat and homogeneous Universe. Therefore, the metric is given by the FLRW metric
\begin{equation}
    ds^2 = -dt^2 + a^2(t) \delta_{ij} dx^i dx^j,
\end{equation}
with this particular form of the metric, the non-metricity scalar Eq.~(\ref{eqn:Q}) in the coincident gauge is given by $Q = 6H^2$ with $H = \dot{a}/a$ the Hubble factor. \\

Since we have set a connection and a metric, we can in principle derive the generalized Friedmann equations by substituting our assumptions in the field equations Eq.~(\ref{eqn:raisedFieldEquations}). However, we still need to know the specific form of the tensor $\Theta^\mu_{\;\;\nu}$. Combining the definition \newline $\Theta_{\mu \nu} = g^{\alpha \beta} \dfrac{\delta T_{\alpha \beta}}{\delta g^{\mu \nu}}$ with equation Eq.~(\ref{eqn:stress-energyGeneral}) gives us
\begin{equation}
    \label{eqn:ThetaTensor}
    \Theta_{\mu\nu} = \mathcal{L}_m g_{\mu \nu} - 2 T_{\mu\nu}.
\end{equation}

As we can see, $\Theta_{\mu\nu}$ depends on the matter Lagrangian. Since there are two possibilities of the matter Lagrangian that can derive the stress-energy tensor of a perfect fluid Eq.~(\ref{eqn:stress-energyPerfectFluid}), we have a degeneracy on the form of the $\Theta^\mu_{\;\;\nu} $ tensor that will have an impact on the generalized Friedmann equations. We will study how this degeneracy affects this equations and the cosmological implications that it has. Let us start with the matter Lagrangian $\mathcal{L}_m = p$ with $p$ the pressure.

\subsection{Generalized Friedmann equations for $\mathcal{L}_m = p$}

By taking $\mathcal{L}_m = p$, the theta tensor is given by $\Theta_{\mu\nu} = p g_{\mu\nu} - 2T_{\mu\nu}$. Now, by using the FLRW metric from the field equations we can find that the generalized Friedmann equations are given by
\begin{equation}
\label{eqn:firstFriedmannEquation-p}
    \frac{f}{2} - 6 f_Q H^2 = 8\pi \rho + f_T (\rho+p),
\end{equation}
and
\begin{equation}
\label{eqn:secondFriedmannEquation-p}
  -8\pi p +2(  \dot{H}f_Q + \dot{f_Q} H) = \frac{f}{2} -6f_QH^2,
\end{equation}
using Eq.~(\ref{eqn:firstFriedmannEquation-p}) in Eq.~(\ref{eqn:secondFriedmannEquation-p}) we get

\begin{equation}
\label{eqn:secondFriedmannEquationfinal-p}
    \dot{H}f_Q + \dot{f_Q} H = 4\pi(\rho+p) + \frac{f_T}{2} (\rho+p),
\end{equation}
where $H = \dot{a}/a$ is the Hubble factor and the dots represent derivatives with respect to cosmic time. As we can see, the first equation Eq.~(\ref{eqn:firstFriedmannEquation-p}) has an additional term $f_T (\rho + p)$ with a coupling between $\rho$ and $p$. Moreover, the function $f(Q,T) = -\dfrac{Q+2\Lambda}{G}$ recovers the $\Lambda$CDM model with $G$ being the Newton constant.

\subsection{Generalized Friedmann equations for $\mathcal{L}_m = -\rho$}

Let us now consider the case $\mathcal{L}_m = -\rho$. Consequently the $\Theta$ tensor is $\Theta_{\mu\nu} = -\rho g_{\mu\nu} - 2T_{\mu\nu}$. Then, in this case the generalized Friedmann equations are given by
\begin{equation}
\label{eqn:firstFriedmannEquation-rho}
    \frac{f}{2} - 6f_Q H^2 = 8\pi \rho,
\end{equation}
and
\begin{equation}
\label{eqn:secondFriedmannEquation-rho}
   \frac{f}{2}-2(\dot{f_Q} H+f_Q( \dot{H} +3H^2))  = -8\pi p - f_T (\rho + p).
\end{equation}
using Eq.~(\ref{eqn:firstFriedmannEquation-rho}) in Eq.~(\ref{eqn:secondFriedmannEquation-rho}) we get
\begin{equation}
\label{eqn:secondFriedmannEquationfinal-rho}
    \dot{H} f_Q + \dot{f_Q} H = 4\pi (\rho+p) + \frac{f_T}{2} (\rho + p).
\end{equation}

While the second Friedmann equations Eq.~(\ref{eqn:secondFriedmannEquationfinal-rho}) and Eq.~(\ref{eqn:secondFriedmannEquationfinal-p}) are identical for this model, the first Friedmann equations Eq.~(\ref{eqn:firstFriedmannEquation-p}) and Eq.~(\ref{eqn:firstFriedmannEquation-rho}) differ by 
a coupling between $\rho$ and $p$. 
Moreover, with this matter Lagrangian, the first Friedmann equation is identical to the $f(Q)$ case \cite{jimenez2020cosmology}. 

\section{Effects of the Matter Lagrangian Degeneracy in the density evolution}

\label{sec:Effects 6}

We will now turn to the study of the effects of the matter Lagrangian degeneracy on the density evolution. To do so, we need to take the Levi-Civita covariant derivative of the field equations Eq.~(\ref{eqn:raisedFieldEquations}) \cite{xu2019f}
\begin{align}
    &\mathcal{D}_\mu \left( f_T (T^\mu_{\;\;\nu} + \Theta^\mu_{\;\;\nu}) - 8\pi T^\mu_{\;\;\nu} \right) - \frac{1}{2} f_T \partial_\nu T = \nonumber \\
    &\frac{1}{\sqrt{-g}} Q_\mu \nabla_\alpha \left(f_Q \sqrt{-g} P^{\alpha \mu}_{\;\;\;\;\nu} \right) - \frac{8\pi}{\sqrt{-g}} \nabla_\alpha \nabla_\mu H_\nu^{\;\;\alpha\mu},
\end{align}
where $\nabla$ is the total connection (equal to the standard derivative in the coincident gauge) and $\mathcal{D}$ is the Levi-Civita connection. We can use the connection field equation Eq.~(\ref{eqn:connectionFieldEquations}) to rewrite this as
\begin{align}
\label{eqn:stress-energy-evolution}
    &\mathcal{D}_\mu \left( f_T (T^\mu_{\;\;\nu} + \Theta^\mu_{\;\;\nu}) - 8\pi T^\mu_{\;\;\nu} \right) - \frac{1}{2} f_T \partial_\nu T = \nonumber \\
    &\frac{1}{\sqrt{-g}} Q_\mu \nabla_\alpha \left(f_Q \sqrt{-g} P^{\alpha \mu}_{\;\;\;\;\nu} \right) + \frac{2}{\sqrt{-g}} \nabla_\alpha \nabla_\mu (f_Q \sqrt{-g} P^{\alpha \mu}_{\;\;\;\;\nu}),
\end{align}
since $\Theta^\mu_{\;\;\nu}$ appears explicitly in equation Eq.~(\ref{eqn:stress-energy-evolution}), the form of the matter Lagrangian will influence the results. We will focus on the zeroth component of this equation since that will give us the equation for the evolution of density. 

\subsection{Density evolution for $\mathcal{L}_m = p$}

By taking the coincident gauge and the matter Lagrangian $\mathcal{L}_m = p$ in the zeroth component of equation Eq.~(\ref{eqn:stress-energy-evolution}), we get
\begin{equation}
\label{eqn:matter-density-p}
    \dot{\rho} = - \dfrac{3H (f_T + 8\pi)  \rho(1+w)}{8\pi + \dfrac{1}{2}f_T(3-c_s^2) - f_{TT} \rho (1+w) (1-3c_s^2)},
\end{equation}
where $w \equiv p/\rho$, $f_{TT} \equiv d^2f/dT^2 = df_T/dT$ and $c_s^2 \equiv \dot{p}/\dot{\rho}$. If we set $f_T = 0$, we recover the standard case $\dot{\rho} + 3H\rho(1+w) = 0$. However, the coupling of $T$ in the action Lagrangian causes the density evolution to have additional contributions. This can be interpreted as stress-energy transfer between matter and geometry and particle production/annihilation \cite{xu2019f,wu2018palatini}. \\ 
\subsection{Density evolution for $\mathcal{L}_m = -\rho$}

Now, let us study the density evolution for the case with $\mathcal{L}_m = -\rho$. If we introduce this form for the matter Lagrangian in the zeroth component of equation Eq.~(\ref{eqn:stress-energy-evolution}), we get
\begin{equation}
\label{eqn:matter-density-rho}
    \dot{\rho} = -\dfrac{3H(f_T+8\pi)\rho(1+w)}{8\pi+\dfrac{1}{2}f_T (1-3c_s^2)},
\end{equation}
then, in this case, we do not have the presence of the second derivative $f_{TT}$. This suggests that solving the equation for density should be easier in this case. In addition to this, the $f_T$ term in the denominator differs from the $\mathcal{L}_m = p$ case. It is also interesting to note that if we consider radiation $w = 1/3 = c_s^2$, then
\begin{equation}
    \dot{\rho_{r}} = -4H \left( 1 + \frac{f_T}{8\pi} \right) \rho_r.
\end{equation}

Therefore, in the radiation case, the evolution simplifies considerably. Furthermore, if $f_T << 8\pi$, we recover the standard case $\dot{\rho_r} = -4H\rho_r$. This also suggests that it is much easier to work with this Lagrangian since the equation for the density evolution reduces considerably. 

\section{Specific Cosmological Models}
\label{sec:Specific Cosmological Models 7}

In order to see the cosmological effects that a different matter Lagrangian has, we will consider some cosmological models. We will begin with the simple $f(Q,T) = f_1(Q) + 16\pi a T$ function, where $a$ is a parameter and $f_1(Q)$ is an arbitrary function of the non-metricity scalar $Q$. We will also work with a function of the form $f(Q,T)= f_1(Q) - 8\pi aT^2$. It is useful to consider this function because it exhibits a quadratic dependence on the trace of the stress-energy tensor and it has been shown that a model of this form challenges the $\Lambda$CDM standard model in a background perspective \cite{najera2021fitting}. And finally, we will consider a new kind of $f(Q,T)$ function with a logarithmic dependence on the trace of the stress-energy tensor. This function will be of the form $f(Q,T) =f_1(Q) + 16\pi a \ln \left(b \dfrac{T}{T_0} \right)$ where $a$ and $b$ are parameters and $T_0$ is the trace of the stress-energy tensor today. The general procedure to follow will be:
\begin{enumerate}
    \item To derive the density evolution $\rho$ for matter and radiation and compare the results. 
    \item To consider the simple case with $f_1(Q) = - \dfrac{Q + 2\Lambda}{G}$, where $\Lambda$ and $G$ are the cosmological and Newton constants respectively. These functions will reduce to the $\Lambda$CDM case when we turn off the dependence of the $f(Q,T)$ function on $T$. 
    \item To derive the generalized Friedmann equations with this form of $f_1(Q)$.
    \item To use cosmic clocks data to perform a Monte Carlo Markov Chain (MCMC) method and get the confidence regions of the parameter vector. This will be done to compare the performance between the two matter Lagrangians. 
\end{enumerate}

This procedure will be illustrative to test the effects that each Lagrangian has on the density evolution, the generalized Friedmann equations and on an MCMC analysis. However, it has to be noted that these simple models cannot solve the cosmological constant problem since they include it. Then, the functions to be considered will be
\begin{enumerate}
    \item $f(Q,T) = - \dfrac{Q+2\Lambda}{G} + 16\pi a T$
    \item $f(Q,T) = - \left( \dfrac{Q+2\Lambda}{G} + 8\pi a T^2 \right)$
    \item $f(Q,T) = - \dfrac{Q+2\Lambda}{G} + 16\pi a \ln \left( b \dfrac{T}{T_0} \right)$
\end{enumerate}

\subsection{Density evolution}

The first step to study the physical effects that the matter Lagrangian degeneracy has in $f(Q,T)$ gravity is to see how the density evolution changes with redshift. We will compute the results for both forms of the matter Lagrangian. In each case, we will derive the evolution of dust ($w=0$) and radiation ($w=1/3$) remembering that  $a=1/(1+z)$. 

\subsubsection{Density evolution for $f(Q,T) = - \dfrac{Q+2\Lambda}{G} + 16\pi a T$ and $\mathcal{L}_m = p$}

This model is the simplest extension into $f(Q,T)$ gravity and it reduces to the $\Lambda$CDM case when $a=0$. If we consider $\mathcal{L}_m = p$ as the matter Lagrangian, we need to take equation Eq.~(\ref{eqn:matter-density-p}). In this case $f_T = 16\pi a$ and $f_{TT} = 0$, therefore
 \begin{equation}
     \rho(z) = \rho_0 (1+z)^\alpha,
 \end{equation}
 where $\alpha = \dfrac{3(1+2a)(1+w)}{1+a(3-c_s^2)}$. In the matter case $w = 0$ ($c_s^2 = 0$)  then
 \begin{equation}
     \rho_M(z) = \rho_{0M} (1+z)^{\frac{3(1+2a)}{1+3a}},
 \end{equation}
 and it reduces to the standard case when $a \to 0$. Now, for radiation $w=1/3$ ($c_s^2 = 1/3$) and then
 \begin{equation}
     \rho_{r}(z) = \rho_{0r} (1+z)^{\frac{12(1+2a)}{3+8a}},
 \end{equation}
 which also reduces to the standard case when $a \to 0$.
 
 \subsubsection{Density evolution for $f(Q,T) = - \dfrac{Q+2\Lambda}{G} + 16\pi a T$ and $\mathcal{L}_m = -\rho$}
 
 If we now consider $\mathcal{L}_m = -\rho$ as the matter Lagrangian, the density evolution is given by
 \begin{equation}
     \rho(z) = \rho_0 (1+z)^\alpha,
 \end{equation}
 where $\alpha = \dfrac{3(1+2a)(1+w)}{1+a(1-3c_s^2)}$. The particular cases for matter and radiation are
 \begin{equation}
     \rho_M(z) = \rho_{0M} (1+z)^{\frac{3(1+2a)}{1+a}},
 \end{equation}
 and
 \begin{equation}
     \rho_r(z) = \rho_{0r} (1+z)^{4(1+2a)},
 \end{equation}
 and they reduce to the standard case when $a\to0$. \\
 
 As we can see, the difference between both Lagrangians resides in the denominators of the exponents of $1+z$. In the matter case with $\mathcal{L}_m = p$, we have $1+3a$ while for $\mathcal{L}_m = -\rho$, it is $1+a$. This difference can have effects when fitting the models to data as we will see later. The same happens in the case of radiation with a change in the denominator from $3+8a$ for $\mathcal{L}_m = p$ to $1$ for $\mathcal{L}_m = -\rho$.
 
 \subsubsection{Density evolution for $f(Q,T) = - \left( \dfrac{Q+2\Lambda}{G} + 8\pi a T^2 \right)$ and $\mathcal{L}_m=p$}
 
 This kind of function has a quadratic dependance on $T$. It also reduces to $\Lambda$CDM when $a=0$. However, when $a\neq 0$, it can have interesting features as we will see. Since we will start with $\mathcal{L}_m = p$,
 \begin{equation}
    f_T=-16\pi aT=16\pi a\rho(1-3w),
\end{equation}
and
\begin{equation}
    f_{TT} = -16\pi a,
\end{equation}
then with the aid of equation Eq.~(\ref{eqn:matter-density-p}) follows that for matter ($w=0$)
\begin{equation}
    \rho_{M}^2(z) (2a\rho_M(z)+1)^3 - \rho^2_{0M} (2a\rho_{0M}+1)^3 (1+z)^6 = 0,
\end{equation}
which is a fifth-order differential equation. Therefore, it cannot be in general solved analytically and if we want to solve it, we must perform a numerical method. If we divide this solution by $\rho^2_{0M}$ and also set $a = a'/\rho_{0M}$ (this is a helpful change of variable as we will see in the next subsection) 
\begin{equation}
    \label{eqn:densityEvolutionMatterFifthOrder}
    \left(\frac{\rho_M(z)}{\rho_{0M}}\right)^2 \left( 2a' \dfrac{\rho_M(z)}{\rho_{0M}} + 1 \right)^3 - (2a' + 1)^3 (1+z)^6 = 0.
\end{equation}

This equation enables us to solve for $\rho_M/\rho_{0M}$ which will be helpful when solving the modified Friedmann equation. For the case of radiation ($w=1/3$)
\begin{equation}
    \rho_{r}(z) = \rho_{0r} (1+z)^4,
\end{equation}
which is identical to the standard case. While the matter case presented a fifth-order differential equation, it is interesting to note that for radiation, the expression is unchanged. 

\subsubsection{Density evolution for $f(Q,T) = - \left( \dfrac{Q+2\Lambda}{G} + 8\pi a T^2 \right)$ and $\mathcal{L}_m=-\rho$}

Let us now consider the alternative form of the matter Lagrangian ($\mathcal{L}_m = -\rho$). For the matter case ($w=0$)
\begin{align}
    &\rho_M(z) = \frac{(1+z)^3}{1+2a\rho_{0M}} \biggl[ a\rho_{0M}^2 (1+z)^3  \nonumber \\
    &+ \sqrt{a^2 \rho_{0M}^4 (1+z)^6+ \rho_{0M}^2 (1+2a\rho_{0M})}\biggr] ,
\end{align}
which is the solution to a quadratic equation and reduces to the standard case for $a=0$. Let us define $a=a'/\rho_{0M}$, then this result can be rewriten as
\begin{equation}
    \label{eqn:densityEvolutionMatterSecondOrder}
    \frac{\rho_M(z)}{\rho_{0M}} = \frac{(1+z)^3}{1+2a'} \left[ a' (1+z)^3 + \sqrt{a'^2 (1+z)^6+ (1+2a')} \right],
\end{equation}
which enables us to get the adimensional result $\rho_M/\rho_{0M}$. This result is interesting since this form of the Lagrangian gives an analytic solution to the density of matter as a function of redshift. This fact suggests that it is easier to work with this matter Lagrangian. However, we still need to see the performance when fitting the model with data. We will do this later on the paper. We can also perceive that the matter density includes a contribution proportional to $(1+z)^3$ however, it includes two higher-order $(1+z)^6$ contributions that come from the coupling of $T^2$ in the action Lagrangian. \\

When considering radiation (w=1/3)
\begin{equation}
    \rho_{r}(z) = \rho_{0r} (1+z)^4,
\end{equation}
which is identical to the standard case. As we can see, for both Lagrangians, radiation evolves identically.

\subsubsection{Density evolution for $f(Q,T) = - \dfrac{Q+2\Lambda}{G} + 16\pi a \ln \left( b \dfrac{T}{T_0} \right)$ and $\mathcal{L}_m=p$ or $\mathcal{L}_m = -\rho$}

This model is an interesting one since it exhibits the property of having the same density evolution for the matter case independently of the election of $\mathcal{L}_m$. If we equal equations Eq.~(\ref{eqn:matter-density-p}) and Eq.~(\ref{eqn:matter-density-rho}), we get
\begin{equation}
    \frac{1}{2} f_T (1-3c_s^2) = \frac{1}{2} f_T (3-c_s^2) - f_{TT} \rho (1+w) (1-3c_s^2),
\end{equation}
by taking the dust case ($w=0$), we can solve this equation as
\begin{equation}
    f(Q,T) = f_1(Q) + 16\pi a \ln \left(b \frac{T}{T_0} \right),
\end{equation}
which is a logarithmic model. Therefore, for this model the evolution of matter will be identical for both forms of the matter Lagrangian. By taking this logarithmic model into equation Eq.~(\ref{eqn:matter-density-rho})
\begin{equation}
    \rho_M(z) = a + \sqrt{a^2 + \rho_{0M} (\rho_{0M} - 2a) (1+z)^6}.
\end{equation}

This result shows a dependence proportional to $(1+z)^6$ plus a linear dependence that disappears when $a=0$. Although the density evolution is the same for $\mathcal{L}_m = p$ and $\mathcal{L}_m = -\rho$, the generalized Friedmann equations should not be equal. We will see whether this is the case, and if it is different how it affects the performance of an MCMC. This result can be simplified by setting $a=a' \rho_{0M}$ and then
\begin{equation}
    \label{eqn:densityLogarithmic}
    \frac{\rho_M(z)}{\rho_{0M}} = a' + \sqrt{a'^2+(1-2a')(1+z)^6}.
\end{equation}

In the case of radiation, we have encountered a problem while studying this model. Since radiation has an equation of state of $w=1/3$, it has a null trace of the stress-energy tensor. This implies that in $T/T_0$ we have a division between two zeros, which is physically impossible. Hence, this model would be unable to reproduce the whole story of the Universe since early times are dominated by radiation, a fluid that is incompatible with this logarithmic function. However, we can still take it as a toy model to study the effects that different Lagrangians can have in $f(Q,T)$ gravity.

\subsection{Generalized Friedmann equations}

Now that we have derived the density evolution for dust matter and radiation, we can focus on computing the generalized Friedmann equations using this results and equation Eq.~(\ref{eqn:firstFriedmannEquation-p}) for the case of $\mathcal{L}_m = p$ and equation Eq.~(\ref{eqn:firstFriedmannEquation-rho}) for $\mathcal{L}_m = -\rho$. This is neeeded in order to test our models with observational data.
 
 \subsubsection{Modified Friedmann equation for $f(Q,T) = - \dfrac{Q+2\Lambda}{G} + 16\pi a T$ and $\mathcal{L}_m = p$}
 
 Let us compute the first Friedmann equation Eq.~(\ref{eqn:firstFriedmannEquation-p})
 \begin{equation}
     3H^2 = \Lambda + G\rho \left[ 8\pi + 8\pi a (3-w) \right],
 \end{equation}
 if we set the definition
 \begin{equation}
     \Omega_i = \frac{8\pi G \rho_{0i}}{3H_0^2} \left( 1 + (3-w_i) a \right),
 \end{equation}
 where the $i$-th index represents the matter and radiation fluids and $\Lambda = 3H_0^2 \Omega_\Lambda$, then the modified Friedmann equation becomes
 \begin{equation}
     H(z) = H_0 \sqrt{\Omega_\Lambda + \Omega_r (1+z)^\beta + \Omega_M (1+z)^\gamma},
 \end{equation}
 where $\beta = \dfrac{12(1+2a)}{3+8a}$ and $\gamma = \dfrac{3(1+2a)}{1+3a}$. As we can see
 \begin{equation}
     \Omega_\Lambda + \Omega_r + \Omega_M = 1,
 \end{equation}
 and therefore this model has four independent parameters, one more than the $\Lambda$CDM model.
 
 \subsubsection{Modified Friedmann equation for $f(Q,T) = - \dfrac{Q+2\Lambda}{G} + 16\pi a T$ and $\mathcal{L}_m = -\rho$}
 
 In this case, the first Friedmann equation (\ref{eqn:firstFriedmannEquation-rho}) is 
 \begin{equation}
     3H^2 = \Lambda + G\rho \left[ 8\pi + 8\pi(1-3w)a \right],
 \end{equation}
 and if we define
 \begin{equation}
     \Omega_i = \frac{8\pi G \rho_{0i}}{3H_0^2} (1+(1-3w_i)a),
 \end{equation}
 the equation is
 \begin{equation}
     H(z) = H_0 \sqrt{\Omega_\Lambda + \Omega_r (1+z)^\beta + \Omega_M (1+z)^\gamma},
 \end{equation}
 where $\beta= 4(1+2a)$ and $\gamma = \dfrac{3(1+2a)}{1+a}$. By setting $z=0$, we also recover the relation
 \begin{equation}
     \Omega_\Lambda + \Omega_r + \Omega_M = 1,
 \end{equation}
 and hence the model has four independent components. The difference between the Hubble factors reside in the exponents of the radiation and matter parts. This differences might play an important role in a numerical analysis.
 
 \subsubsection{Modified Friedmann equation for $f(Q,T) = - \left( \dfrac{Q+2\Lambda}{G} + 8\pi a T^2 \right)$ and $\mathcal{L}_m = p$}
 
 In this case, the first modified Friedmann equation gives us Eq.~(\ref{eqn:firstFriedmannEquation-p})
 \begin{equation}
     3H^2 = \Lambda + 8\pi G \rho_r + 8\pi G \rho_{M} \left( 1+ \frac{5}{2} a \rho_{M} \right),
 \end{equation}
 and if we define $a = a'/\rho_{0M}$ and also
 \begin{equation}
     \Omega_i = \frac{8\pi G \rho_{0i}}{3H_0^2} \left(1+\frac{1}{2} a' (1-3w_i) (5+w_i)\right),
 \end{equation}
 where $i$ represents matter and radiation, then the Hubble factor is 
 \begin{equation}
     \left(\frac{H(z)}{H_0}\right)^2 = \Omega_\Lambda + \Omega_M \frac{\rho_M}{\rho_{0M}} \left( \frac{2+5a' \dfrac{\rho_M}{\rho_{0M}}}{2+5a'} \right) + \Omega_r (1+z)^4,
 \end{equation}
 where $\dfrac{\rho_M}{\rho_{0M}}$ is given by solving equation Eq.~(\ref{eqn:densityEvolutionMatterFifthOrder}). By setting $z=0$, we recover the closure relation
 \begin{equation}
     \Omega_\Lambda + \Omega_M + \Omega_r = 1,
 \end{equation}
 and hence the model has four independent components. 
 
 \subsubsection{Modified Friedmann equation for $f(Q,T) = - \left( \dfrac{Q+2\Lambda}{G} + 8\pi a T^2 \right)$ and $\mathcal{L}_m = -\rho$}
 
 For this matter Lagrangian, the generalized Friedmann equation Eq.~(\ref{eqn:firstFriedmannEquation-rho}) is 
 \begin{equation}
     3H^2 = \Lambda + 8\pi G \rho_r + 8\pi G \rho_M \left( 1 + \frac{a}{2} \rho_M \right),
 \end{equation}
 if we define $a = a'/\rho_{0M}$ and
 \begin{equation}
     \Omega_i = \frac{8\pi G \rho_{0i}}{3H_0^2} \left( 1 + \frac{1}{2} a' (1-3w)^2 \right),
 \end{equation}
 the Hubble factor is given by
 \begin{equation}
     \left( \frac{H(z)}{H_0} \right)^2 = \Omega_\Lambda + \Omega_M \frac{\rho_M}{\rho_{0M}} \left( \dfrac{2+a'\dfrac{\rho_M}{\rho_{0M}}}{2+a'} \right) + \Omega_r (1+z)^4, 
 \end{equation}
 and we can get $\rho_M/\rho_{0M}$ with the quadratic solution Eq.~(\ref{eqn:densityEvolutionMatterSecondOrder}). The generalized Friedmann equations only differ in the coefficient of $a' \rho_M/\rho_{0M}$ (5 for $\mathcal{L}_m = p$ and 1 for $\mathcal{L}_m = -\rho$). However, the solution of the density evolution is much easier for this alternative Lagrangian since it presents a simple quadratic equation while $\mathcal{L}_m = p$ has a fifth order differential equation. This fact gives an advantage to the $\mathcal{L}_m = -\rho$ case that will ease the numerical computations in the MCMC.
 
 \subsubsection{Modified Friedmann equation for $f(Q,T) = - \dfrac{Q+2\Lambda}{G} + 16\pi a \ln \left( b \dfrac{T}{T_0} \right)$ and $\mathcal{L}_m = p$}

For this function, the evolution of dust matter was invariant with a change of matter Lagrangian. However, radiation was not compatible since it presented a quotient between two zeros and also $f_T \to \infty$ for radiation. Hence, this model is incompatible with a radiation-dominated Universe and because of this, it is physically impossible. However, we will study it to see how the matter Lagrangian degeneracy affects the results in $f(Q,T)$ gravity. We will assume that the Universe is composed of dust matter without radiation to avoid the problems that this model has. The first modified Friedmann equation Eq.~(\ref{eqn:firstFriedmannEquation-p}) is
\begin{equation}
    3H^2 = \Lambda + 8\pi G \rho_M - 16\pi G a - 8\pi G a \ln \left( b \frac{\rho_M}{\rho_{0M}} \right),
\end{equation}
we can define $a = a' \rho_{0M}$ and also 
\begin{equation}
    \Omega_M = \frac{8\pi G \rho_{0M}}{3H_0^2} (1-2a'),
\end{equation}
and then the Hubble factor is
\begin{equation}
    \left( \frac{H(z)}{H_0} \right)^2 = \Omega_\Lambda + \frac{\Omega_M}{1-2a'} \left[ \frac{\rho_M}{\rho_{0M}} - a' \left( 2 + \ln \left( b \frac{\rho_{M}}{\rho_{0M}} \right) \right) \right],
\end{equation}
with $\rho_M/\rho_{0M}$ given by equation Eq.~(\ref{eqn:densityLogarithmic}). If we set $z=0$, we recover the relation
\begin{equation}
    \Omega_\Lambda + \Omega_M = 1,
\end{equation}
and the model has three independent components (remember that we are not considering radiation). The result shows the logarithmic dependence plus a constant term $-(2a'\Omega_M)/(1-2a')$ that can act as a contribution to the cosmological constant. It is not a cosmological constant, but a couple constant between geometry and matter. This is an interesting feature since this contribution can reproduce a cosmological constant-like evolution. However, the model has the serious problem of not being compatible with radiation. 

\subsubsection{Modified Friedmann equation for $f(Q,T) = - \dfrac{Q+2\Lambda}{G} + 16\pi a \ln \left( b \dfrac{T}{T_0} \right)$ and $\mathcal{L}_m = -\rho$}

The modified Friedmann equation (\ref{eqn:firstFriedmannEquation-rho}) is
\begin{equation}
    3H^2 = \Lambda + 8\pi G \rho_M - 8\pi G a \ln \left( b \frac{\rho_M}{\rho_{0M}} \right),
\end{equation}
and if we set the definitions $a = a' \rho_{0M}$ and
\begin{equation}
    \Omega_M = \frac{8\pi G \rho_{0M}}{3H_0^2},
\end{equation}
then
\begin{equation}
    \left( \frac{H(z)}{H_0} \right)^2 = \Omega_\Lambda + \Omega_M \left[ \frac{\rho_M}{\rho_{0M}} - a' \ln \left( b \frac{\rho_M}{\rho_{0M}} \right) \right],
\end{equation}
which shows a logarithmic contribution. However, in this case, we do not have the cosmological constant-like evolution. This is because the generalized Friedmann equation Eq.~(\ref{eqn:firstFriedmannEquation-rho}) does not include the $f_T (\rho+p)$ term as Eq.~(\ref{eqn:firstFriedmannEquation-p}) does. The solution of $\rho_M/\rho_{0M}$ is given by equation Eq.~(\ref{eqn:densityLogarithmic}) and the closure relation also holds here 
\begin{equation}
    \Omega_\Lambda + \Omega_M = 0,
\end{equation}
which confirms that this model has three independent parameters. 

\subsection{Numerical analysis}

Until now, we have derived the effects that the matter Lagrangian degeneracy has on the generalized Friedmann equations and density evolution. We have seen that the sole fact of changing the Lagrangian can have huge changes in the results. In this subsection, we will see how this degeneracy can cause the cosmological constraints to differ. For that purpose, we will use the cosmic clocks 2016 compilation \cite{moresco20166}.

\subsubsection{Cosmic clocks 2016}

This compilation \cite{moresco20166} presents 30 measurements of the Hubble factor $H(z)$ along with their uncertainties $\sigma_{H(z)}$ at different redshift $z$. These measurements are model independent. From the definitions of the scale factor, the Hubble factor and redshift we can write $H(z) = -\dfrac{1}{1+z} \dfrac{dz}{dt}$. Then, from the differential age evolution of two early-type galaxies we can measure the ratio $\dfrac{dz}{dt}$ \cite{jimenez2002constraining} and with that the Hubble factor can be computed in a model independent way. In order to compare the observational data with the model, we will compute the $\chi^2$ function given by
\begin{equation}
    \chi^2 = \sum_{n=1}^{30} \left( \frac{H(\Theta,z_n)-H_\text{obs}(z_n)}{\sigma_\text{obs}(z_n)} \right)^2,
\end{equation}
where $\Theta = \{ H_0, \Omega_M, a, b \}$ represents the parameter vector, $z_n$ is the redshift of the $n$-th sample, $H_\text{obs}(z_n)$ and $\sigma_\text{obs}(z_n)$ the observational Hubble factor and uncertainty of the Hubble factor respectively. The quantity $H(\Theta,z_n)$ is the theoretical Hubble factor for a given point in the parameter space. The best value of this parameter vector will be given by minimizing the $\chi^2$ function. For all our models, the $\Omega_\Lambda$ parameter can be written in terms of $\Omega_M$ and $\Omega_r$ and we therefore did not include it in the analysis. Also, since at late times, radiation is neglibible, we will consider $\Omega_r = 0$. \\

After getting the best value of the parameter vector, we will perform a Monte Carlo Markov Chain (MCMC) method with the \texttt{python} package \texttt{emcee} \cite{foreman2013emcee}. We will initialize 32 random walkers in a tiny Gaussian ball around the values of the parameters given by the minimization of the $\chi^2$ function. We will run the MCMC method until the chains are 50 times the auto-correlation time (the number of steps required to have an independent set of chains) as suggested by the software \cite{foreman2013emcee}. These chains can be saved in H5 files and then analyzed with the \texttt{GetDist} software that plots the 68.3\% and 95\% confidence regions and also gets the best fit MCMC value with the 68.3\% confidence level. We will repeat this analysis for both matter Lagrangians ($\mathcal{L}_m = p$ and $\mathcal{L}_m = -\rho$) and for the three models considered. For the quadratic model $f(Q,T) = - \left( \dfrac{Q+2\Lambda}{G} + 8\pi a T^2 \right)$ we will use the bisection method to solve the numerical equation Eq.~(\ref{eqn:densityEvolutionMatterFifthOrder}) and with that the corresponding Hubble factor. 

\begin{table*}
\caption{\label{tab:results} Median values and and 68.3\% uncertainty of the cosmological parameters for the $f(Q,T)$ functions and both matter Lagrangians considered and the cosmological parameters $\Theta = \{ H_0, \Omega_M, a, b \}$. \textit{Note: For the quadratic and logarithmic models $a$ is the $a'$ parameter previously discussed.}}
\begin{ruledtabular}
\begin{tabular}{cccccc}
 Model & $\mathcal{L}_m$ & $H_0$ (km/s/Mpc) & $\Omega_M$ & $a$ ($a'$) & $b$ \\ \hline
 \multirow{2}{*}{$- \dfrac{Q+2\Lambda}{G} + 16\pi a T$} & $p$ & $65.3\pm 3.5$ & $0.58\pm 0.18$ & $0.46^{+0.46}_{-0.33}$ & - \\
    & $-\rho$ & $67.9\pm 3.9$ & $0.39^{+0.11}_{-0.30}$ & $0.056^{+0.087}_{-0.29}$ & - \\ \hline
    \multirow{2}{*}{$- \left( \dfrac{Q+2\Lambda}{G} + 8\pi a T^2 \right)$} & $p$ & $65.6\pm 3.3$ & $0.476^{+0.086}_{-0.12}$ & $0.25\pm 0.14$ & - \\
    & $ -\rho $ & $66.3^{+4.3}_{-3.9}$ & $0.418^{+0.096}_{-0.19}$ & $-0.008^{+0.014}_{-0.012}$ & - \\ \hline
    \multirow{2}{*}{$-\dfrac{Q+2\Lambda}{G} + 16\pi a \ln \left( b \dfrac{T}{T_0} \right)$} & $p$ & $63.7\pm 4.6$ & $0.55^{+0.18}_{-0.20}$ & $-0.89^{+0.48}_{-0.96}$ & $1.43\pm 0.36$ \\
    &$-\rho$ & $66\pm 8$ & $0.328^{+0.068}_{-0.17}$ & $-0.31^{+0.70}_{-0.19}$ & $1.42\pm 0.36$ \\
\end{tabular}
\end{ruledtabular}
\end{table*}

\subsubsection{Analysis}

\begin{figure}
    \centering
    \includegraphics{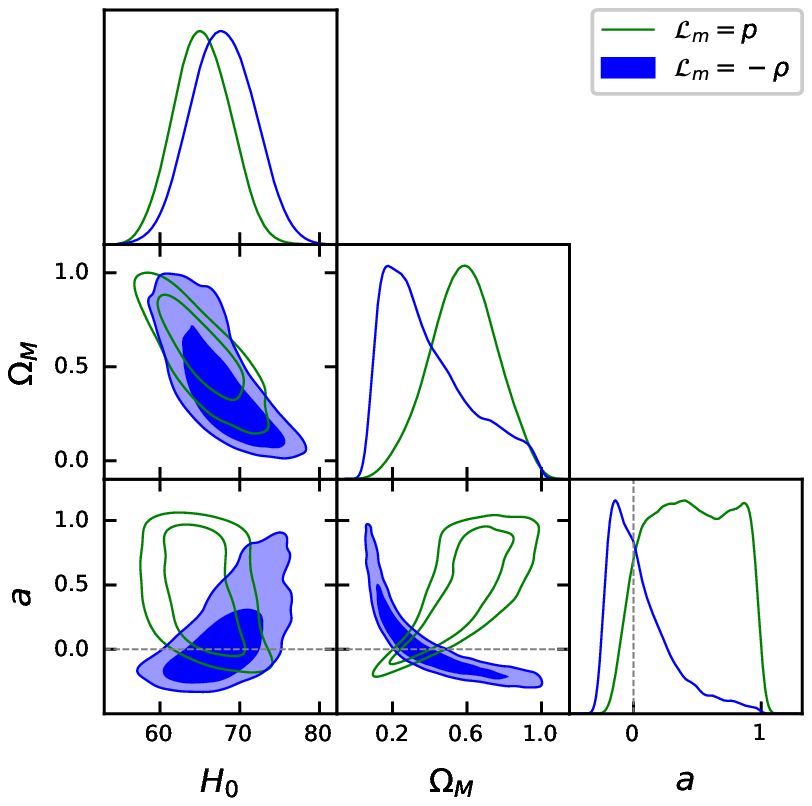}
    \caption{68.3\% and 95\% confidence regions of the parameters $\Theta = \{ H_0, \Omega_M, a \}$ and the cosmic clocks 2016 compilation \cite{moresco20166} for the linear $f(Q,T) = - \dfrac{Q+2\Lambda}{G} + 16\pi a T$ model and both matter Lagrangians $\mathcal{L}_m = p$ and $\mathcal{L}_m = -\rho$. The dotted lines represent the $\Lambda$CDM limit.}
    \label{fig:linear_model}
\end{figure}

\begin{figure}
    \centering
    \includegraphics{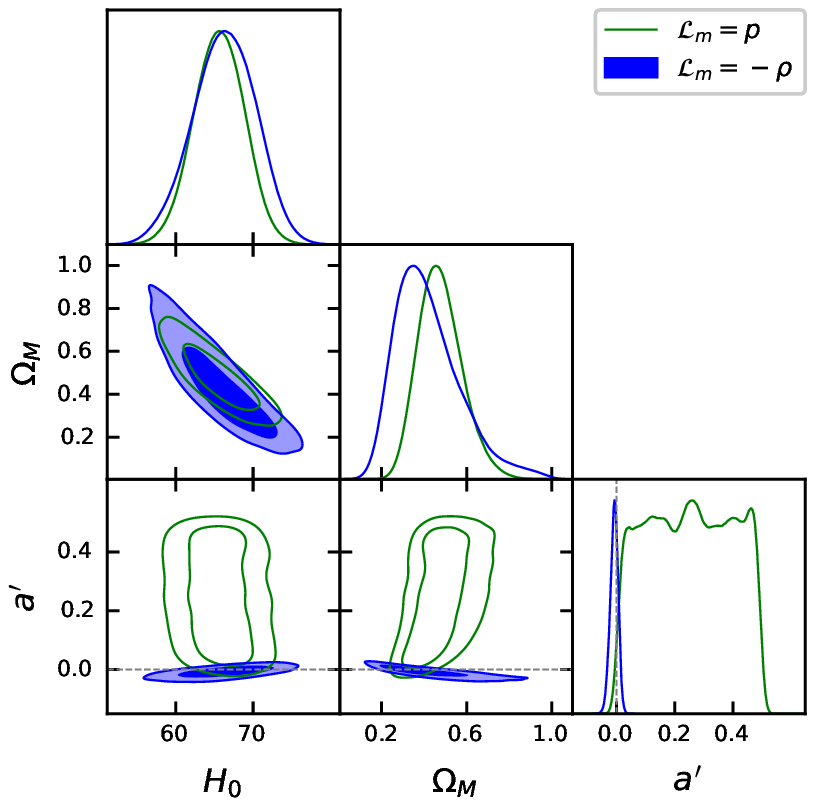}
    \caption{68.3\% and 95\% confidence regions of the parameters $\Theta = \{ H_0, \Omega_M, a' \}$ and the cosmic clocks 2016 compilation \cite{moresco20166} for the quadratic $f(Q,T) = - \left( \dfrac{Q+2\Lambda}{G} + 8\pi a T^2 \right)$ model and both matter Lagrangians $\mathcal{L}_m = p$ and $\mathcal{L}_m = -\rho$. The dotted lines represent the $\Lambda$CDM limit.}
    \label{fig:quadratic_model}
\end{figure}

\begin{figure}
    \centering
    \includegraphics{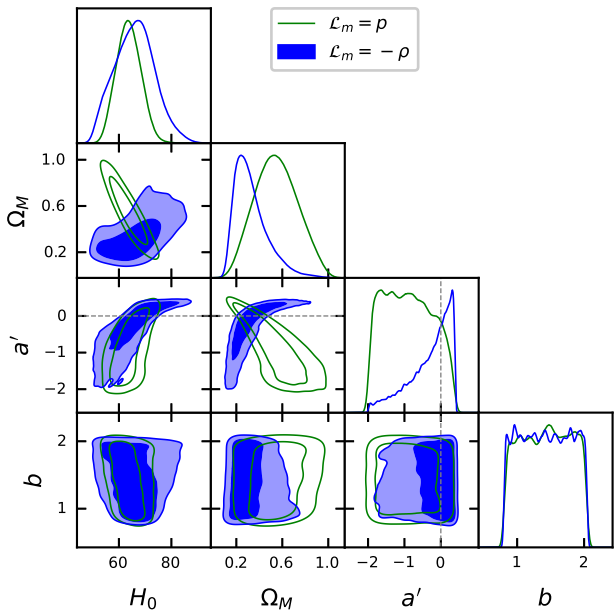}
    \caption{68.3\% and 95\% confidence regions of the parameters $\Theta = \{ H_0, \Omega_M, a', b \}$ and the cosmic clocks 2016 compilation \cite{moresco20166} for the logarithmic $f(Q,T) = - \dfrac{Q+2\Lambda}{G} + 16\pi a \ln \left( b \dfrac{T}{T_0} \right)$ model and both matter Lagrangians $\mathcal{L}_m = p$ and $\mathcal{L}_m = -\rho$. The dotted lines represent the $\Lambda$CDM limit.}
    \label{fig:logarithmic_model}
\end{figure}

We have presented the results of the median and 68.3\% confidence intervals in table \ref{tab:results}. Furthermore, the 68.3\% and 95\% confidence contours are presented in figure \ref{fig:linear_model} for $f(Q,T) = - \dfrac{Q+2\Lambda}{G} + 16\pi a T$, figure \ref{fig:quadratic_model} for $f(Q,T) = - \left( \dfrac{Q+2\Lambda}{G} + 8\pi a T^2 \right)$ and figure \ref{fig:logarithmic_model} for $f(Q,T) = - \dfrac{Q+2\Lambda}{G} + 16\pi a \ln \left( b \dfrac{T}{T_0} \right)$. For the quadratic and logarithmic models, we plotted the $a'$ parameter instead of $a$. As it can be seen in these plots, the $\Lambda$CDM limit ($a=0$) lies within the 95\% confidence regions for all cases. \\

In addition to this, we can see that the median, standard deviations and confidence regions differ significantly when changing the matter Lagrangian. We need to note that we are taking the same $f(Q,T)$ function and we are only changing the Lagrangian. For the linear model, figure \ref{fig:linear_model} shows that the matter content of the Universe when considering $\mathcal{L}_m = -\rho$ is lower than with $\mathcal{L}_m = p$. Furthermore, the $a$ parameter has a more restricted contour with $\mathcal{L}_m = -\rho$. For the quadratic model, the plots are similar for both Lagrangians and the $H_0$ and $\Omega_M$ parameters. However, for the $a'$ parameter, the matter Lagrangian $\mathcal{L}_m = -\rho$ shows a more restricted contour and a Gaussian-like form while for $\mathcal{L}_m = p$ the contour did not converge. In addition to this, when performing the MCMC method, this model presented numerical problems in the range $a'<-0.05$ for $\mathcal{L}_m = p$. This behavior might be due to the numerical equation required to compute the density Eq.~(\ref{eqn:densityEvolutionMatterFifthOrder}). For the logarithmic model, figure \ref{fig:logarithmic_model} shows a clear difference in the results for $\Omega_M$ and $a'$. The matter quantity was much higher for $\mathcal{L}_m = p$ than for $\mathcal{L}_m = -\rho$. For the $a'$ parameter, the contours have totally different forms and both did not converge. Finally, the $b$ parameter did not converge in both cases. \\

These results show that changing the matter Lagrangian can have important changes in the median and uncertainty values as well on the confidence regions when performing an MCMC method. Therefore, the sole action of changing the matter Lagrangian has a huge impact on the Cosmological parameters in $f(Q,T)$ gravity. This impact occurred with a small catalog of 30 measurements of the Hubble factor. It can be even bigger when considering bigger catalogs as Type 1a Supernovas, Gamma Ray Bursts or Gravitational Waves. 

\subsection{General analysis}

We have computed the density evolution equations for dust matter and radiation, the generalized Friedmann equations and performed MCMC methods to get the best values and confidence regions of three $f(Q,T)$ gravity functions. We repeated this computations for two alternate matter Lagrangians $\mathcal{L}_m = p$ and $\mathcal{L}_m = -\rho$ that give rise to the same form of the stress-energy tensor of a perfect fluid Eq.~(\ref{eqn:stress-energyPerfectFluid}). \\

We have seen that the sole fact of this matter Lagrangian degeneracy can have huge mathematical and cosmological implications. It was particularly interesting to see that for a quadratic model $f(Q,T) = - \left( \dfrac{Q+2\Lambda}{G} + 8\pi a T^2 \right)$ the density equation for dust matter is a fifth order equation for $\mathcal{L}_m = p$ Eq.~(\ref{eqn:densityEvolutionMatterFifthOrder}) while a second order one for $\mathcal{L}_m = -\rho$ Eq.~(\ref{eqn:densityEvolutionMatterSecondOrder}). This happens for every quadratic model on $T$. This fact eased the numerical MCMC method for the latter case while it presented numerical problems for the former. Another interesting consequence was that the median and uncertainty values along with the confidence regions of the $\Omega_M$ and the modified $a$ (or $a'$) and $b$ parameters can change significantly when considering an alternate form of the matter Lagrangian. This happened with a catalog of only 30 observational events. This impact can be significantly higher when considering bigger catalogs. Therefore, every theoretical or observational study in $f(Q,T)$ should consider both forms of the matter Lagrangian to account for this mathematical and observational changes. 
 
\section{Conclusions}
\label{sec:Conclusions}

In the present paper, we have investigated the theoretical and cosmological effects of the matter Lagrangian degeneracy in $f (Q, T )$ gravity  
where $Q$ is the non-metricity scalar and $T$ is the trace of the stress-energy tensor, which is a modified theory of gravity in the symmetric teleparallel framework. Since both $\mathcal{L}_m = p$ and $\mathcal{L}_m = -\rho$ matter Lagrangians can give the same form of the stress-energy tensor of a perfect fluid Eq.~(\ref{eqn:stress-energyPerfectFluid}), there will be a degeneracy on the $f(Q,T)$ gravity field equations Eq.~(\ref{eqn:raisedFieldEquations}) as the $\Theta_{\mu\nu} = g^{\alpha \beta} \delta T_{\alpha \beta}/\delta g^{\mu \nu}$ depends on the matter Lagrangian.

The degeneracy has consequences on the density evolution of dust matter and radiation, on the form of the generalized Friedmann equations and also changes in the cosmological parameters when performing MCMC analyses. For $\mathcal{L}_m = p$, the first Friedmann equation (Eq.~(\ref{eqn:firstFriedmannEquation-p})) includes a coupling between $\rho$ and $p$ with an additional term $f_T (\rho + p)$. This term does not appear for the $\mathcal{L}_m = -\rho$ case (Eq.~(\ref{eqn:firstFriedmannEquation-rho})). For the density evolution equation $d\rho/dt$, the difference resides on the form of the denominator. For $\mathcal{L}_m = p$ a dependence on $f_T = df/dT$ and $f_{TT} = d^2 f/dT^2$ appears (Eq.~(\ref{eqn:matter-density-p})) while $\mathcal{L}_m = -\rho$ only has a dependance of $f_T = df/dT$ (Eq.~(\ref{eqn:matter-density-rho})). 

To deeply study the effects that the matter Lagrangian degeneracy has in $f(Q,T)$ gravity, we derived the particular solutions for three $f(Q,T)$ functions. One linear $f(Q,T) = - \dfrac{Q+2\Lambda}{G} + 16\pi a T$, one quadratic $f(Q,T) = - \dfrac{Q+2\Lambda}{G} + 16\pi a \ln \left( b \dfrac{T}{T_0} \right)$ and a logarithmic one $f(Q,T) = - \left( \dfrac{Q+2\Lambda}{G} + 8\pi a T^2 \right)$. The density evolution of dust matter ($w=0$) was different in the three models and for the quadratic one, the matter Lagrangian $\mathcal{L}_m = p$ gave a fifth-order linear Eq.~(\ref{eqn:densityEvolutionMatterFifthOrder}) while $\mathcal{L}_m = -\rho$ gave a simple quadratic equation much easier to solve. Radiation ($w=1/3$) had a different density evolution for the linear model but an standard evolution for the quadratic one $\rho_r(z) = \rho_{r0} (1+z)^4$. In addition to this, the logarithmic function presented serious issues with the radiation evolution, having an infinite $f_T$. This fact makes this logarithmic model incompatible with a radiation dominated Universe. 

We used the results of the density evolution, and we derived the generalized Friedmann equations to compute the theoretical value of the Hubble parameter $H(z)$. With it, we performed Monte Carlo Markov Chain methods to get the median, uncertainties and confidence regions of the cosmological parameters in order to study the cosmological effects that the matter Lagrangian degeneracy has on $f(Q,T)$ gravity. We used the cosmic clocks 2016 compilation \cite{moresco20166}, a catalog with 30 measurements of the Hubble factor at different redshift. Our results indicated us that by changing the matter Lagrangian, the confidence regions can change considerably, particularly the ones corresponding to $\Omega_M$, the matter content, and $a$ and $b$, the modified gravity parameters. Also, for the quadratic model, the value of $a'$ is much more constrainted with $\mathcal{L}_m=-\rho$ than with $\mathcal{L}_m = p$. This happened because the simple quadratic evolution of dust matter eased the numerical method for the former case while it presented numerical problems and non-convergence of the parameter for the latter one.

Due to all the differences that we have encountered when changing the matter Lagrangian, future theoretical and cosmological studies regarding $f(Q,T)$ gravity should consider both forms of the matter Lagrangian ($\mathcal{L}_m = p$ and $\mathcal{L}_m = -\rho$) to account for the mathematical and observational differences and to potentially infer which election of the matter Lagrangian presents an advantage over the other when considering several cosmological models and over the free parameters. 


\nocite{*}

\bibliography{apssamp}

\end{document}